# `imaka - a ground-layer adaptive optics system on Maunakea

Mark R. Chun[a*], Olivier Lai[b], Douglas Toomey[c], Jessica R. Lu[d], Max Service[a], Christoph Baranec[a], Simon Thibault[e], Denis Brousseau[e], Yutaka Hayano[f], Shin Oya[f], Shane Santi[g], Christopher Kingery[h], Keith Loss[h], John Gardiner[i], Brad Steele[i]

[a]Institute for Astronomy, University of Hawaii-Manoa, Hilo, Hi 96720; [b]Observatoire de la Côte d'Azur, Nice, France; [c]Mauna Kea Infrared LLC, Hilo, HI 96720; [d]Institute for Astronomy, University of Hawaii-Manoa, Honolulu, Hi 96822; [e]Universite Laval, 2375, rue de la Terrasse, local 2104 Québec (Québec) G1V 0A6; [f]Subaru Telescope, National Astronomical Observatory of Japan, Hilo, HI 96720; [g]Dream Telescopes & Acc., Inc., [h]RockWest Composite, San Diego, CA 92173, [i]Quartus Engineering, San Diego, CA.

## ABSTRACT

We present the integration status for `imaka, the ground-layer adaptive optics (GLAO) system on the University of Hawaii 2.2-meter telescope on Maunakea, Hawaii. This wide-field GLAO pathfinder system exploits Maunakea's highly confined ground layer and weak free-atmosphere to push the corrected field of view to ~1/3 of a degree, an areal field approaching an order of magnitude larger than any existing or planned GLAO system, with a FWHM ~ 0.33" in the visible and near infrared. We discuss the unique design aspects of the instrument, the driving science cases and how they impact the system, and how we will demonstrate these cases on the sky.

## 1. INTRODUCTION

Many of the second-generation astronomical adaptive optics systems are pushing toward larger corrected fields of view. Multi-conjugate adaptive optics (MCAO) systems such as Gemini's GEMS[1], ESO's MAD demonstrator [2], and the solar MCAO systems at the Dunn [3], GREGOR [4], and Big Bear [5] solar telescopes seek to achieve corrected images with modest Strehl ratios over roughly one arcminute fields of view. Multi-object adaptive optics (MOAO) demonstrator systems such as CANARY [6] and Raven [7] look to extend these corrections to several small "single-object" fields distributed within a field of a few arcminutes. These systems recognize that the optical turbulence that gives rise to the optical aberrations is distributed throughout the Earth's atmosphere and that a simple "single-conjugate" adaptive optics (SCAO) system[1] measures and applies a correction specific to a particular path through the atmosphere (e.g. in the direction of the wavefront sensor guide star). As a result, the quality of the correction varies across the field with the highest quality portion of the field limited to fields of a few arcsecond at visible wavelengths and a few tens of arcseconds in the near-infrared. For many astronomical observations the objects or the collection of objects extend over a much larger field of view. Another approach to extend the corrected field of view is to selectively correct for just the aberrations that arise close to the ground. These ground-layer adaptive optics (GLAO) systems trade the amount of image correction for an increase in the corrected field of view and a more uniform image quality [8].

Above the summit of Maunakea in Hawaii, the distribution of optical turbulence appears ideal for ground-layer adaptive optics. The optical turbulence close to the ground is highly confined within the first tens of meters above the ground [9][10] and the optical turbulence in the free-atmosphere is weak [9][11]. At older telescopes such as the University of Hawaii 2.2-meter telescope, 2/3rds of the total optical turbulence arises within the first 30 meters above the ground. The combination of the thin ground layer and weak turbulence in the upper atmosphere presents the possibility of achieving angular resolutions a factor of 2-3 better than delivered naturally by the atmosphere over fields of view several tens of arcminutes in diameter. This areal field of view is more than an order of magnitude larger than previously envisioned for GLAO. Recent results from the RAVEN MOAO demonstrator at the Subaru telescope on Maunakea indicate that with a GLAO correction, image resolutions between 0.1" and 0.3", improvements of a factor of 3 over natural seeing, can be obtained at near infrared wavelengths over fields of view of 1-2 arcminutes [12].

---

1     To be explicit, by SCAO we refer to adaptive optics systems with a single correcting element conjugate to the ground <u>and</u> a single wavefront sensor.

*mchun@ifa.hawaii.edu; phone 1 808 932-2317; fax 1 808 933-0737;

`imaka is a project to demonstrate GLAO and its scientific gains on Maunakea with corrected fields of view up to 0.3×0.4 degrees at the University of Hawaii 2.2-meter (UH88") telescope [13]. Our simulations [14] suggest we can reach the free-atmosphere seeing (~1/3" in the visible) over these fields of view. The project will feed design studies for GLAO systems on the larger telescopes on Maunakea including the ULTIMATE-Subaru project [15] and Keck telescopes.

## 2. `IMAKA DESIGN

The primary goals of `imaka are to demonstrate the performance of GLAO on large fields of view from the visible to near-infrared and to demonstrate the gains realized for GLAO science cases. Table 1 lists our key objectives. For each of the cases we have identified fields we will observe to demonstrate each of the cases. The cases are not exhaustive but include fields to demonstrate the gains in resolution/field as well as our ability to extract quantitative measurements of position, flux, and morphology from astronomical observations with GLAO systems.

Table 1: GLAO Technical and Science Demonstrations for `imaka

| Item | Measurement | Comments |
| --- | --- | --- |
| TD1 | Image Quality | Quantify the full-width half-maximum and the equivalent noise area as a function of wavelength, optical turbulence profiles, and guide star asterism (brightness and separations). Compare with model predictions. |
| TD2 | PSF Stability | Measure the spatial and temporal variation of the point-spread function and compare with model predictions and measured optical turbulence profiles |
| TD3 | Astrometric Stability | Measure the position repeatability and stability of stars within the field over periods of minutes to nights. |
| SD1 | Astrometry | Extract differential-astrometric measurements for stellar objects over a wide field. |
| SD2 | Photometry | Extract quantitative flux measurements from crowded stellar fields |
| SD3 | Morphology | Extract morphologies of resolved objects in very long exposure observations. |

We note that in each of the above cases the implementation of the instrument can affect our ability to quantify the performance. In some cases the science drivers have technical requirements that may be beyond the budget of the instrument. For example, for the science case of the discovery of exoplanet around other stars via their astrometric signal requires relative position measurements of signals less than one milliarcsecond on the sky. At the final plate scale of `imaka this is equivalent to a position variation of about 150 nm. This puts strong constraints on the stability of both the optics and opto-mechanics of the instrument. A variation of the mirror figure at the level of tens of nanometers, due to say a changing gravity vector as the telescope changes observing elevation angle, could on its own introduce variations in the focal plane positions of a star that are appreciable to the signal we're trying to measure. For these cases, the observations are designed (along with calibration procedures) to minimize and characterize the instrumental contributions in order to quantify any limitation imposed by GLAO.

A principal requirement in the above demonstrations is that the optical design and optics of an adaptive optics system deliver good image quality at focal planes and at its pupil-planes. For Maunakea, excellent free-atmosphere seeing (a proxy for the GLAO corrected image) can be as good as a full-width half-maximum of FWHM~0.25" at visible wavelengths so at a minimum the optical design should not degrade this image quality. In addition we note that, unlike narrow-field AO systems, the deformable mirror in a GLAO system will only correct for static optical aberrations on optics conjugate to altitudes close to the ground so the optical systems figure error should also not degrade the image quality. The second key optical requirement for a GLAO system is the fidelity of the pupil image on the deformable mirror across the science field of view. Any pupil distortion across the field, equivalent to a misregistration of the DM across the field, must be small compared to the actuator pitch to ensure that the GLAO correction is applied uniformly across the field. In addition, we seek a design that minimizes the number of surfaces, allows for the procurement of low figure error mirrors, and if there is an optical relay, favors larger deformable mirrors with small tilt angles on all powered elements. The ideal case is to implement a GLAO system using an adaptive secondary or primary mirror.

However, given the cost of such a system, we were forced for `imaka to find an optical relay design but that would still demonstrate GLAO correction over fields of view of tens of arcminutes.

The typical approach, using two or more off-axis parabolas, does not deliver the required image quality or pupil quality for fields larger than a few arcminutes. The optical design of the `imaka (Figure 1 below) is based on a near-infrared camera for the MMT GLAO by Baranec et al [16] and uses a "broken" Offner relay where the input and output focal ratios differ (f/10 in and f/13.25 out) and the pupil is before the second element. This design has several important features. First, the pupil position, normally at the secondary of the Offner relay, is moved away from the convex secondary to a location where a flat deformable mirror can be placed. Second, the design offers a useful trade of the image quality for the field of view. For `imaka our goal is to push GLAO to as large a field as possible so we settled on a total field of view of 24' x 18' but with the image quality optimized over the central 11' x 11' (science). The central "science" field of view matches the largest optical camera we envision working with the instrument, while the larger outer portion of the field provides sufficient image quality and field of view to find natural guide stars for the Shack-Hartmann wavefront sensors. Over the central 11' x 11' the design FWHM is better than 0.15" which increases the expected best case GLAO performance by about 0.05". This is a non-negligible degradation of the delivered images but one we were willing to make in exchange for being able to obtain guide stars out as far as 12' off axis. By driving the GLAO correction with guide stars at the edges of the field, we can probe GLAO on fields of view significantly larger than the fields (<3' diameters) of previous systems such as the MMT GLAO system [17][18], SOAR SAM [19], LBT ARGOS, [20], Raven [7] and MAD [2] experiments. Third, the Offner relay is tolerant to misalignment with the strictest tolerances on the tilt error on the two concave powered elements (~1 arcminute). The key here is that the aberrations introduced by the first concave powered element are minimized by the tilt of the second concave element. Recognizing this eases the opto-mechanical requirements of the structure and simplifies the alignment procedure. Finally, all the powered elements in the design have spherical surfaces, which simplifies the procurement of the large (~400mm diameter) mirrors and simplifies aligning the optics (e.g. a decenter error of the mirror can be compensated with a piston and tilt of the mirror).

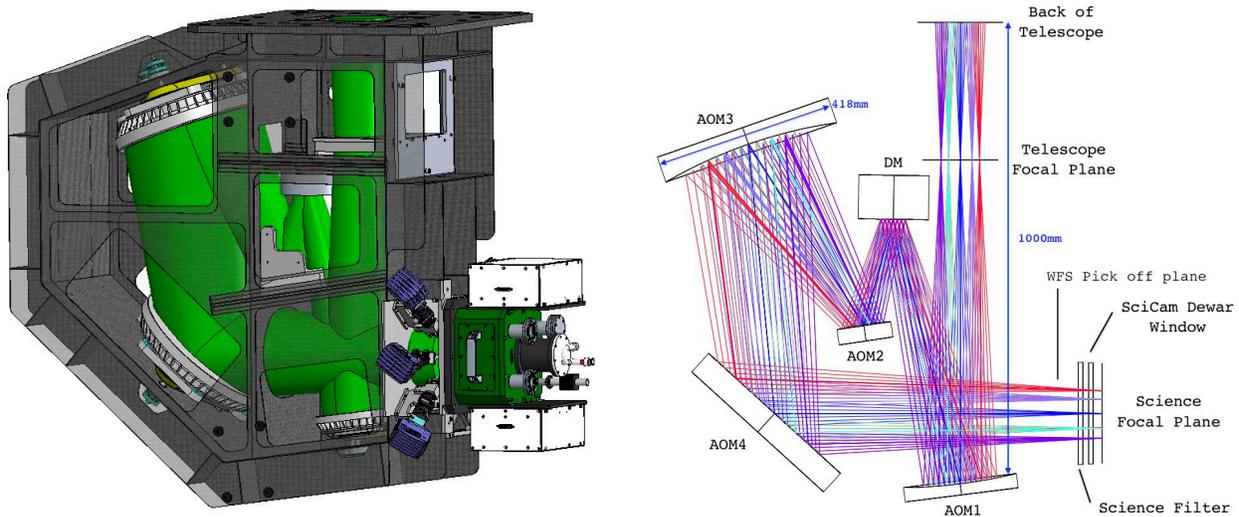

Figure 1: CAD renderings of the opto-mechanical layout of `imaka (left) and the optical layout (right). Light from the telescope (shown in green) enters from the top of the image thru the telescope interface plate. The telescope f/10 Cassegrain focus resides approximately 300mm below this surface. The first sphere reimages the telescope primary mirror onto the deformable mirror while the next two powered elements (a convex and a concave sphere produce a f/13.3 final focus. The final large element in the system is a large flat 400mm in diameter. It folds the optical path and keeps the overall structure dimensions about 1m x 1m x 0.5m. The envelope of optical beams in green shows the full 0.3 x 0.4 degree field of view. The system's cameras, wavefront sensor in purple and the science camera in green and white, are shown at the lower right. The shelf at the top right resides at the telescope focal plane and provides space for calibration sources. In the figure on the right, we note that the distance from the back of the telescope to the first optical element is about one meter.

To make use of the corrected science focal plane, we will use the UH88" facility 10kpixel x 10kpixel CCD camera based on a DALSA STA1600 monolithic detector and a near-infrared camera based on the first science-grade Teledyne H4RG-

15 detector [21]. These cameras will provide fields of view of 11'x11' in the visible (0.4-1.0 microns) and 7'x7' in the near-infrared (1-1.8um) respectively. Both of the cameras are compatible with the exit focal plane of `imaka but only one camera will work on the instrument at a time. The UH88" STA1600 camera is under commissioning at the telescope while the first science grade H4RG-15 detectors are undergoing lab/on-sky testing [21].

Since `imaka uses natural guide stars and, importantly, has a modest budget, we've made some important concessions on the instrument to simplify and reduce cost. In particular, the wavefront sensors will pick off light from stars in the outer parts of the field and will be manually positioned for each guide star asterism. The instrument will be configured at sea level for each run - mounting holes for the wavefront sensors will be machined in the WFS mounting plate prior to the first observation for each asterism of guide stars and matching calibration sources will be placed in the entrance focus calibration unit in order to align the wavefront sensors and generate interaction matrices for the WFSs and DM. However, on the telescope a change in target will require manual intervention to place the WFSs at the appropriate locations. This limits our observing efficiency and our ability to observe many different targets/guide star asterisms. This limitation is most important for (1) the direct comparison of the delivered GLAO performance as a function of asterism shapes/sizes and (2) in dithering the field across the detector for science observations either to build a larger field of view or to remove cosmetic defects or sky background variations. In the first case, we plan to always compare the performance against the measured optical turbulence profiler (a MASS/DIMM unit that operates nightly on Maunakea) and our simulations [14]. At a practical level, there are, in any case a limited number of fields we can target that have a sufficiently bright (R<11) constellation of guide stars. In the second case, the wavefront sensor subapertures have a field of view of 4 arcseconds. This ensures linearity of the gradient measurement over the dynamic range required for GLAO measurements but also provides for the possibility to make small dithers by introducing offsets to the "null" position of the wavefront sensors. These small dithers help remove the effects of bad pixels in the science camera but the dithers are not large enough to mosaic a larger science field or to extract the sky background in the near-infrared. As a result, our operational model for the instrument is that of a testbed rather than a facility instrument.

## 3. IMAKA IMPLEMENTATION CHALLENGES

The choice of an optical relay for `imaka has several implementation challenges. As noted above, in order to obtain good image quality and good pupil quality at the deformable mirror, the design pushes to large optics that are closely packed. This results in an instrument that is large (~one meter in length) and potentially much heavier than typical instruments on the UH88" telescope. We estimated that the optics (largest are ~420mm in diameter) fabricated from solid glass substrates along with aluminum mounts would exceed 100kg and the enclosure if made from aluminum would be close to 150 kg. Together these two pieces of the system would exceed our allowable mass budget of 230 kg.

### 3.1. Mirrors and mounts

For the large mirrors we implemented engineered lightweight cast Borofloat33® mirrors with carbon-fiber whiffletree mounts (9-pt for the larger mirrors). Dream Telescopes and Accessories Inc. fabricated the two large concave spheres (AOM1 and AOM3), the large fold flat mirror (AOM4), as well as their mounts. Dream Telescopes also polished the two concave spheres and sub-contracted the polishing of the large flat. The three mirrors were tested and delivered attached to the whiffletree mount. The delivered mass of the large flat (AOM4) with its 9-point carbon-fiber whiffletree mount (Figure 2) is 6.2 kg. This is an estimated weight savings of around 70% over more conventional approaches.

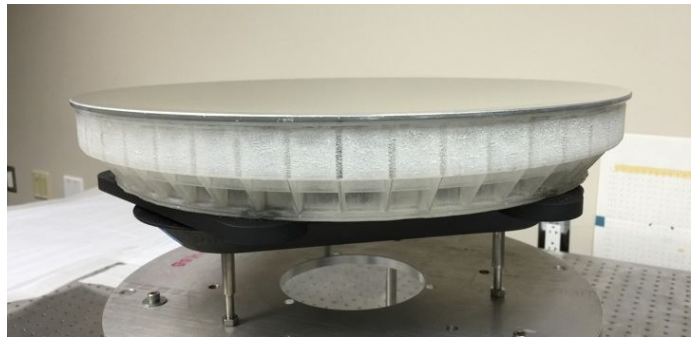

Figure 2. `imaka's large fold flat (AOM4) delivered by Dream Telescopes and Accessories Inc. and mounted in the test stand on an optical bench at the Institute for Astronomy. The lightweight cellular structure of the mirror can be seen as well

as the carbon fiber whiffletree structure (in black) below the mirror. The three studs at the back of the whiffletree mount interface with the `imaka structure via three precision stainless steel balls and a set of races on the `imaka structure (not shown).

### 3.2. Structure/Enclosure

The instrument's mass limit and its large optics also imposed constraints on how we could fabricate the instrument structure/enclosure. The structure had to be light yet stiff to prevent misalignment of the optics with the changing gravity vector the instrument sees at the Cassegrain focus of the telescope. With these operational requirements, we decided against a typical aluminum enclosure and built the structure from panels of carbon fiber laminates. Carbon fiber laminates are lightweight yet stiff and they have a very low coefficient of thermal expansion (CTE) so are dimensionally stable to changes in temperature. However, the choice of carbon fiber also has important implications on the structure's use and the alignment of the optics. In this section we discuss two key issues with its use in astronomical instruments – dimensional changes with moisture content and the absolute precision of the placement of interfaces within the structure.

The `imaka structure, designed by Quartus Engineering and Rock West Composites in San Diego, uses pi/4 quasi-isotropic laminate (autoclave cured) 0.1"-thick panels epoxied together to form the enclosure (Figure 3 below). The structure wall thicknesses are typically between 0.1" to 0.3" (glued up panels) and the top plate of the structure, which interfaces to the telescope, consists of a 0.8"-thick sandwich of eight panels. The CTE of the carbon fiber panels (roughly 3 ppm/deg C) matches that of the mirror glass, which reduces the sensitivity of the instrument to variations in the temperature. The structure's sides, bulkheads, and external "egg-crate" stiffeners interlock with each other via a set of "slotted" joints. The structure panels and aluminum interface bushings are epoxied together in an assembly jig consisting of a large precision steel plate/table with aluminum angle brackets that position the panels and bushings during the assembly process. Ribs, added to the exterior of the enclosure, further stiffen the structure and highlight the flexibility of the design/fabrication. The `imaka enclosure is effectively an inside-out, or exo-skeleton, honeycomb optical bench. The entire carbon fiber enclosure, measuring ~1 meter x 1 meter x 0.5 meter, weighs only 45 kg and does not dominate the mass budget (Table 2).

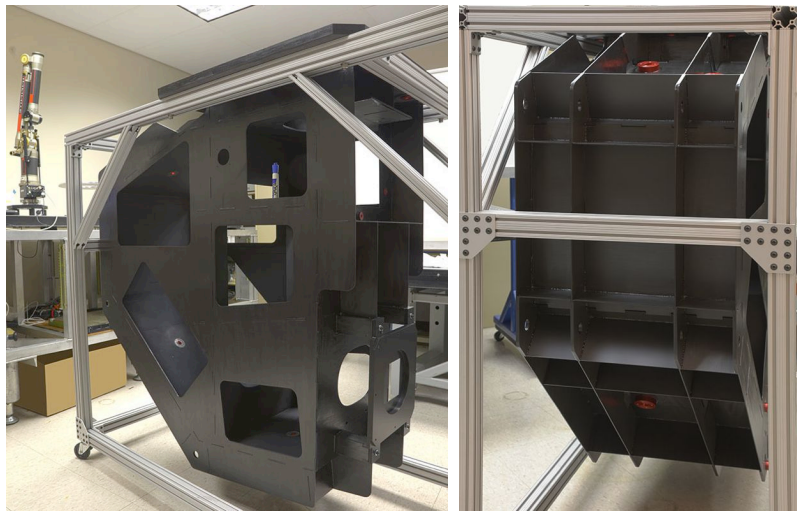

Figure 3. `imaka's carbon-fiber enclosure was designed by Quartus Engineering and Rock West Composites and fabricated by Rock West Composites. The structure is shown unpopulated and mounted in its protective aluminum frame. The final focal plane camera mounts to the aluminum plate at the bottom right of the figure on the left. Note the red aluminum bushings throughout the structure. These are the interfaces between the structure and the optics or calibration unit. The external "egg-crate" construction that stiffens the structure is illustrated in the image on the right. The structure is suspended from its top plate in the aluminum frame. This frame supports the instrument while it is in the lab and protects the structure from accidently impulses.

Table 2. `imaka Mass Budget

| Subsystem | Mass (est) | Comments |
|---|---|---|
| Structure | 45 kg | Structure and aluminum bushings |
| Optics | 20 kg | AOM1-AOM4 and DM |

| Opto-mechanics | 22 kg | Whiffletrees, interfaces to Structure |
| Opto-electronics | 76 kg | WFS/Science cameras, calibration unit |
| Electronics | 50 kg | Computers, DM electronics, power supplies |
| **Total** | **213 kg** | Mass limit 230kg (total for all elements mounted to telescope) |

**Coefficient of Moisture Expansion**

While the structure dimensions are stable to changes in temperature, composite materials are known to change dimension with changes in the moisture content within the laminate [22]. The epoxy resin used within the laminate absorbs and desorbs moisture from the air depending on its exposure and the moisture content in the air. Analogous to the CTE, the coefficient of moisture expansion (CME) describes the strain of the material as a function of the percent weight gain due to the absorbed moisture. After extended exposure to moisture, a composite material will saturate at a moisture content of ~2% so by weight the moisture accounts for only a 2% increase in the total weight of the composite. Expansion/contraction due to the CME is directional. For the quasi-isotropic laminates used here the CME perpendicular to the fibers in the panel is ~5x larger than the CME in the plane of the fibers. However, in the case of `imaka's structure the aspect ratio of the panels is 100-200 so the largest dimensional changes are in the plane of the optics. The total motion in the plane of the roughly one meter side panel is 250 microns from fully saturated (moisture content of 1.5%) to dry (~0.2%) with a $CME_{inplane}$ ~ 200ppm/%moisture. This is a significant change but it is largely within the optical alignment tolerance (~200 microns between elements) and can mostly be compensated by a refocus of the telescope. Note, importantly, that since `imaka's optical design does not use any off-axis optical elements, we are only concerned about the separations in the expansion/contraction. In addition, given that the instrument is being integrated at sea level in Hilo, Hawaii where the humidity is high, the dimensional changes are predictable and can be taken into account in the sea-level alignment.

The important difference between CTE and CME changes are the timescale of the changes. CTE changes occur within a night on timescales of minutes/hours while CME changes in composite materials take days to weeks to change moisture content [22]. Following Shen and Springer [22], we plot in Figure 4 the dimensional changes across a one-meter length panel and thru a 0.2"-thick panel. Half of the dimensional change occurs in the first two weeks assuming that the summit is stable at a relative humidity of 30% (long term average for Maunakea based on data from the Mauna Kea Weather Center between 2000 and 2014). As mentioned above, between each imaka run we will bring the instrument back down to Hilo so during a run (about one week), the CME changes should be much less than our optical alignment tolerance. Eventually `imaka will remain at the summit and we will revisit the optical alignment after several months. Following this the variations due to CME will be about the mean (RH~30%) and are expected to be much smaller.

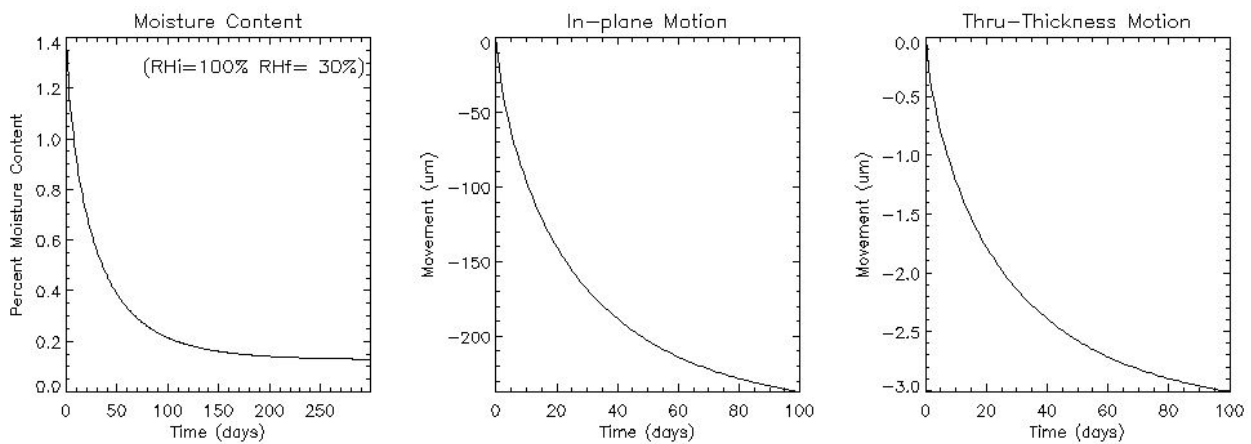

Figure 4. Change in moisture content for a Lamina quasi-isotropic Laminate with a fiber volume of 60%. The plot on the left illustrates the time needed to dry the material from saturated (sea level) to dry (Maunakea). The center and right plots show the in-plane and thru-thickness changes for a one meter long, 0.2" thick panel. Note that half of the motion occurs within the first two weeks.

**Precision of Fabrication**

Our optical alignment tolerances are tightest for the tilts (<1') and the relative center positions (~100 microns) of the powered elements. Our normal procedure for achieving these accuracies is to machine the opto-mechanics from aluminum and front-surface reference the optics in their mounts. For small optics/instruments this procedure positions the optics close and the final alignment can consist of simply shimming the optics in their mounts. In this case, there are two key elements that prevent such an approach. First, the large mirrors are too large to mount in a typical front-surface edge-mount and there was no requirement placed on the precision with which the optic is placed on their mount. The resulting inaccuracies are approximately 0.25-0.5mm. Second, the placement of the individual mirrors within the structure is determined by a set of three aluminum bushings glued into the structure during the structure assembly. The relative position of the three bushings is good (typical machine tolerances) but the absolute position of any set of bushings to any other set is only accurate to ~0.25-0.5mm (e.g. a factor of 5-10 worse). This imprecision is due to errors in the fabrication of the assembly jig, the accuracy of the machining of the panels themselves (e.g. a water jet process), and any motion in the release of the structure from the assembly jig. Alternative fabrication processes, such as machining the optical interfaces into the aluminum bushings after the structure is assembled, might address this but these were beyond the budget of the project.

To achieve the positional accuracies required for the placement of the optics, we needed to measure where the interfaces were in the structure and where the optical surfaces were with respect to their mounts, and we needed a mechanism to adjust the position of the optics with respect to the structure. To do both of these this we (1) measured the position of the interfaces on the structure and the position of the optical surface with respect to its mount using a portable coordinate measuring machine (CMM), and (2) made each optical interface adjustable in position/tilt via shim plates between the optical mounts and structure bushings. We found that the relative position of the interfaces with the structure compared to the CAD model were accurate to better than 0.5mm and typically 0.25mm. A similar process was used to measure the location of the optical surface with respect to its optical mount. For the large optical elements, conical tips were machined into the ends of the three mounting posts that protrude out the back of the whiffletree mount. We used the CMM to measure the location of the front surface (spherical or flat) with respect to the conical holes at the ends of the three posts. In the alignment the CMM will be used to position the optics within the structure in separation and decenter and an alignment telescope plus center targets for each of the optics will define the tilts of each element.

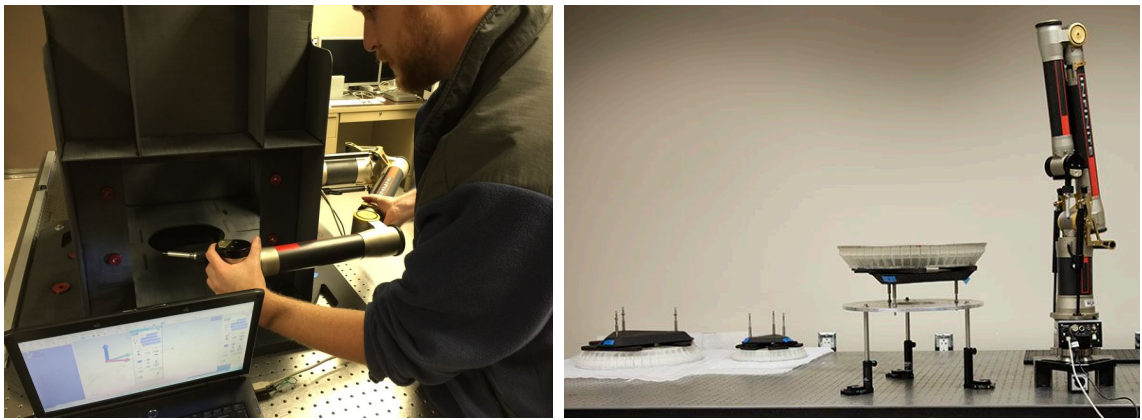

Figure 5. (Left) Measuring the location of the optical interfaces within the structure using a portable coordinate measuring machine builds the precision back into the structure. After measurements, interfaces are known to ~100 microns. Systematic errors in the CMM (e.g. offset drifts with time) limit this accuracy. A similar process to measure the location of the optical system of the three large mirrors with respect to conical holes of the back of each mirror's three mounting posts is done (Right).

In the alignment of `imaka we benefit from the fact that all the powered elements are spheres. A decenter error can be compensated with a tilt and piston of the element – we need to accurately tilt the mirrors but slight offsets in separation and decenter of the on-axis chief ray are not critical. In addition, the Offner relay is very tolerant to errors in the separations between mirrors so the positional requirements should be within reach of the CMM.

## 4. NEXT STEPS

All the optics are in hand and the initial measurements of the structure and mirror/mounts is complete. In the coming weeks we will populate the structure with the optics and proceed thru the optical alignment. On-sky tests with `imaka will begin this fall. We have identified fields for the technical and science demonstrations and expect initial GLAO results in early 2017.

`imaka provides a powerful test-bed for GLAO over very wide fields of view. With its reconfigurable focal planes, we can easily change and deploy different calibration sources or science/wavefront sensor cameras to accommodate different AO experiments. Possible future experiments with the instrument include other wavefront sensing approaches, deployable wavefront sensors, and orthogonal-transfer CCDs for correction of upper-altitude tip/tilt errors. Longer term we plan Rayleigh laser guides stars similar to the Robo-AO system [23][24] to enable general science use.

## ACKNOWLEDGMENTS

The current `imaka work is funded by the National Science Foundation (AST-1310706) and the Mt. Cuba Foundation. C. Baranec acknowledges support from the Alfred P. Sloan Foundation.